\documentclass[aps,prl,showpacs,twocolumn,groupedaddress,amssymb]{revtex4}

\usepackage{graphicx}
\usepackage{dcolumn}
\usepackage{bm}

\begin{document}

\title{Soft x-ray generation based on the two-plasmon decay and the backward Raman scattering in a moderately relativistic electron beam}

\author{S. Son}
\affiliation{169 Snowden Lane, Princeton, NJ, 08540}

\begin{abstract}
A scheme of soft x-ray radiation is proposed. 
In the scheme, 
a moderately intense laser excites plasmons via the two-plasmon 
decay in a relativistic electron beam.
As the second laser encounters the electron beam in the opposite direction, 
it emits soft x-rays via the backward Raman Scattering. 
Our analysis suggests that the effective cross-section of this scattering 
is higher than the Thomson scattering and that  
the conversion efficiency from the pump laser 
to the soft x-ray could be as high as 100\% in the optimal scenario. 
Using the plasmon pump  with  duration of 1-100 pico-second and the electron beam with density of $10^{18} / \mathrm{cc}$ to  $10^{12} / \mathrm{cc}$ and energy of 1-10 MeV, 
soft x-ray of 5 nm to 300 nm with the duration of  10 femto seconds to 
1 pico-secondcan be emitted in the direction of the electron beam.  
Advantages (disadvantages) of the scheme over other schemes are discussed.
\end{abstract}


\maketitle

\section{Introduction}
The soft x-rays, coherent or not, have many important applications, and 
there are increasing needs for powerful and short x-ray source
 in the 
spectroscopy or the semi-conductor lithography\cite{soft, soft2, soft3, litho, litho2, litho3, litho4} and many others.  
 Most of  those applications have not been realized, as the current light  sources need to be improved significantly in the efficiency, the power and the intensity. 

One emerging way to generate intense x-ray  is  to utilize the scattering between an intense laser and a relativistic electron beam, 
 potentially  surpassing the conventional free elecron laser (FEL), the most powerful soft x-ray light so far. 
 The great advances in intense visible-light lasers~\cite{cpa, cpa2, cpa4} and  dense relativistic electron beams~\cite{monoelectron, ebeam} make the technology based on these advances promising;
The laser-based FEL~\cite{Free, Free2}  and non-linear Thomson scattering\cite{thomson} are among them. 

In this paper, the authors  discusses a new approach of soft x-ray generation along this line of thought. 
The relevant physics, that the author takes advantage of, are the two-plasmon decay and the backward Raman scattering (BRS). 
An intense laser, which we call the plasmon pump laser, 
 passes through an electron beam and   
 renders the two-plasmon decay~\cite{liu, langdon} unstable 
 if the wave frequency of the laser is twice of the plasmon frequency 
in the co-moving  frame. 
Among the plasmons excited, 
out interest is the ones whose wave vector is parallel 
to the electron beam direction.
The second laser (the BRS pump laser)  encounters the electron beam in the \textit{opposite} direction and emits soft x-ray (the seed pulse) 
via the BRS with the aforementioned plasmons. 
The frequency of the seed pulse  is given as 

\begin{equation} 
\frac{\omega_{s0}}{\omega_{p0}}  \cong   \frac{ 
\sqrt{(1 + 3 P^2)} + \sqrt{3} \beta P}{ 
{\sqrt{1 + 3 S^{2}} -\sqrt{3} \beta S }} \label{eq:down2} \mathrm{,}
\end{equation}
where $\omega_{s0}$ ($\omega_{p0}$) is the wave frequency of the seed pulse (the BRS pump laser),  $P= S - 1/\sqrt{3} $, and 
$S$ is defined as  
\begin{equation} 
 S =  \left(\frac{\omega_{p0}}{\gamma_0^{1/2}  \omega_{pe}} 
  - \frac{\gamma_0^{1/2}  \omega_{pe}}{\omega_{p0}} \right)/ 2 \sqrt{3}
\mathrm{,}\label{eq:S}
 \nonumber
\end{equation}
where  $\omega_{pe}^2 = 4\pi n_0 e^2 /m_e$ is the plasmon frequency.  Eqs.~(\ref{eq:down2}) and (\ref{eq:S}) will be derived later in this paper. 


According to our analysis, 
the  laser with intensity of $I= 10^{11}\  \mathrm{W} / \mathrm{cm} $ to $I= 10^{13}\  \mathrm{W} / \mathrm{cm}$ excites strong plasmons
to the level of $0.01<\delta n / n_1 <0.5$ in the electron beam 
with the density of $10^{18} / \mathrm{cc} $  to $10^{22} / \mathrm{cc}  $.
The BRS pump with the duration of 1 to 100 pico-seconds could emit 
the soft x-ray of 5 nm to 300 nm 
in the duration of  10 femto-seconds to 1 pico-second
when $2<\gamma_0 < 10 $ and $1<S<5$. 
The effective cross-section of the BRS is much larger 
than the Thomson scattering  and  
the conversion efficiency of the BRS pump laser to the seed pulse 
 could be as high as 100 \%. 
Contrary to the usual BRS, the most of the soft x-ray energy comes 
not from the BRS pump laser but from the plasmon energy;
In this context, the conversion efficiency larger than 100 \% is not contradictory. 

This paper is organized as follows. In Sec.~II, 
we analyze the two-plasmon decay in a moderately relativistic electron beam. 
In Sec.~III, we discuss the BRS scattering and soft x-ray generation deriving Eqs.~(\ref{eq:down2}) and (\ref{eq:S}). 
In Sec.~IV, we estimate the mean-free path of the BRS and the conversion efficiency. We also suggests the physical parameter regime for practical interest.
In Sec.V, we summarize and discuss various aspects of the current scheme.

\section{two-plasmon decay in a moderately relativistic electron beam} 
Let us consider a plasmon pump laser propagating in the same direction 
with the electron beam (density $n_0$ and relativistic factor $\gamma_0$)
 and denote the plasmon pump laser frequency (the wave vector) as  
  $\omega_{pp0}$ ($k_{pp0} $).
It is most convenient to analyze the two-plasmon decay in the co-moving frame
where the electron beam is stationary.  
We denote the co-moving frame (the laboratory frame)  as 1 (0); 
For an example, the wave frequency and the wave vector in the laboratory frame (the co-moving frame) are denoted as 
 $\omega_{pp0}$  and $k_{pp0} $ ($\omega_{pp1}$  and $k_{pp1} $). 
 In the co-moving frame, the electron density is given as 
$n_1 = n_0 / \gamma_0 $ due to the length dilation and 
the plasmon frequency is given as $\omega_3^2 \cong 4 \pi n_0 e^2 /m_e \gamma_0$. 
A photon in the co-moving frame satisfies the dispersion relationship, 
$\omega^2 =  \omega_{pe}^2/\gamma_0 + c^2 k^2 $, 
where $k$ ($\omega$) is the wave vector (frequency) of the photon. 
 The two plasmon decay occurs when $\omega \cong 2\omega_{pe}/\sqrt{\gamma_0}$ or $ ck \cong \sqrt{3} \omega_{pe} / \sqrt{\gamma_0}$.

Consider the case 
when the wave vector of the plasmon pump laser 
is parallel (the same or opposite direction) 
to the electron beam direction.
The wave vectors (wave frequency) of the plasmon pump laser 
between the laboratory frame and the co-moving frame are related as
\begin{eqnarray} 
\omega_{pp0} &=& \gamma_0 \left[ \sqrt{\omega_{pe}^2/\gamma_0 + c^2 k_{pp1}^2 } \pm vk_{pp1} \right] \mathrm{,}  \label{eq:lorentz1} \\  \nonumber \\
k_{pp0} &=&  \gamma_0 \left[ k_{pp1} \pm \frac{\omega_{pp1} }{c}  \frac{v_0}{c} \right] \mathrm{,} \label{eq:lorentz2}
\end{eqnarray}
where $\omega_{pp0}$ and $k_{pp0}$  ($\omega_{pp1}$ and $k_{pp1}$)  
are the wave frequency and the vector of the laser
 in the laboratory frame (the co-moving frame), and the upper case (the lower case) is when the photon wave vector is the same (opposite direction) to the electron beam in the co-moving frame.  
From Eq.~(\ref{eq:lorentz1}) and  
the condition of the two-plasmon decay 
($ck_{pp1}  \cong \sqrt{3} \omega_{pe} / \sqrt{\gamma_0} c$, 
 $\omega_{pp1} \cong 2 \omega_{pe} / \sqrt{\gamma_0}$),  
we obtain 
\begin{eqnarray} 
\omega_{pp0} &\cong& (1 \pm \frac{\sqrt{3}}{2} \beta) \omega_{pp1}  
\mathrm{,} \nonumber \\ \nonumber \\
k_{pp0} &\cong& \left(1 \pm \frac{2}{\sqrt{3}} \beta    \right) k_{pp1} 
 \mathrm{.} \label{eq:ppump} \\ \nonumber  
\end{eqnarray} 
For the upper case (the lower case), the signs of $k_{pp1} $ and $k_{pp0}$
are the same (opposite) and the laser propagates with the same direction (opposite direction) to the beam in the co-moving frame. However, for both cases, 
the laser propagates in the same direction with the beam in the laboratory frame.
For the upper case (lower case), 
the frequency down-shift of the plasmon pump laser 
from the laboratory frame $\omega_{pp0} $ to the co-moving frame $\omega_{pp1} $ is given by the factor $  F=(1 + \sqrt{3} \beta / 2)\gamma_0>1$ 
($F=(1 - \sqrt{3} \beta / 2)\gamma_0$).
For the lower case, assuming $\beta \cong 1$, 
$F > 1 $ ($F<1$) when $\gamma_0 > 7 $ ($\gamma < 7$) so that 
the frequency can be both down-shifted and up-shifted.

When the electron beam density is low, it is better to have a down-shifted laser in the co-moving frame for the two-plasmon frequency condition. 
On the other hand, if 
the electron beam density as high as $10^{20} / \mathrm{cc} $ or higher, 
the down-shifting of the plasmon pump laser renders the two-plasmon decay 
condition not feasible. In this case,  the lower case, which does not down-shift as much as the upper case or even up-shift, is advantageous. 
This is especially the case for the co2 laser. 

Lastly, consider a situation when the plasmon pump laser
in the co-moving frame is propagating to the $\pi/4$ angle to the beam direction. 
This is particular useful 
since the most unstable plasmon would be in the parallel direction 
to the beam~\cite{liu}. 
Denote the wave vector $\mathbf{k}_{pp0} = (k_{pp0x}, 0, k_{pp0z})$ in the laboratory frame. 
From the Lorentz transform, we obtain 
 $k_{pp1x} \cong k_{pp0x} $ and $k_{pp1z} \cong \gamma_0 (1 + 2 \sqrt{2} \beta) k_{pp0z}$ and then  $k_{pp0x} /k_{pp0z} \cong \gamma_0(1+ \sqrt{8} \beta) $ from the condition $k_{pp1x}\cong k_{pp1z}$. 
For a given electron beam, 
the laser should be injected into the beam with the angle given by $k_{pp0x} / k_{pp0z}  \cong \gamma_0 (1+\sqrt{8} \beta) $ and $ck_{pp0x} \cong \omega_{pe} / \sqrt{1.5\gamma_0}$.
The laser direction is almost at the right angle to 
the beam direction and thus the interaction time between the beam and the laser will 
be limited by the spot size of the laser.
%

The density fluctuation due to the plasmon pump laser from the two-plasmon decay 
 is well-analyzed and given as~\cite{liu} 

\begin{eqnarray} 
\left(\frac{\delta n}{n_1 }\right)^2 &\cong& 
  \frac{3}{8 \pi} \left( \frac{c}{v_{te} } \right) \left(\frac{c^2k_{pp1}^2 \gamma_0}{\omega_{pe}^2}\right)
   \left(\frac{e^2E_{pp1}^2 \gamma_0}{m_e^2\omega_{pe}^2 c^2}\right) \nonumber \\ \nonumber \\
 &=& \frac{9}{2\pi} \gamma_0^2\left( \frac{c}{v_{te} } \right)  
\left(\frac{v_q^2 }{c^2}\right) \left(\frac{k_3^2}{k_{pp1}^2}\right)\mathrm{,}  \label{eq:lan} 
  \end{eqnarray}
where $E_{pp1}$ is the electric field strength of the plasmon pump laser in the co-moving frame, 
$v_q = e E_{pp1} / m_e \omega_{ppl}$ is the quiver velocity,  
$k_3 $ is the wave vector of the Langmuir wave, 
$v_{te} $ is the electron thermal velocity in the same frame, and we use 
$ck_{pp1} \cong \sqrt{3} \omega_{pe} / \sqrt{\gamma_0}$ and $\omega_{pp1}= 2ck_{pp1} / \sqrt{3} $.
The threshold condition for the two-plasmon decay is given as $1/3 (v_q/v_{te})^2 k_{pp1} L > 1$~\cite{liu}, where
$L$ is the length scale of the density variation.  
For an example,  for the co2 laser with $k_{pp1} L \cong 100$ and 
the electron beam with the temperature of 1 keV  in the co-moving frame, 
the threshold intensity is given as $I \cong 10^{10}  \ \mathrm{W} / \mathrm{cm}^2 $ for $k_{3} / k_{pp1} \cong 3$. 
One useful fact is that the quiver velocity $v_q $ is invariant under the Lorentz transform if $E_{pp0} \cong B_{pp0} $ and  $E_{pp1} \cong B_{pp1} $, where  $E_{pp0}$ ($B_{pp0}$) is the electric (magnetic) field strength of the plasmon pump laser.
Also note that the kinetic energy spread $\delta E / E $ of the electron beam 
in the laboratory frame is the same order with the velocity spread of the beam in the co-moving frame: 
 $\delta E/ E \cong \delta v / v$. 
Assuming  the beam energy spread in the laboratory frame is between 1 \% and 10 \%, the electron temperature in the co-moving frame is between 50 eV and 5 keV.

\section{soft x-ray generation via the BRS} 

Consider the BRS pump laser propagating 
in the opposite direction to the beam. 
Denote $\omega_{p0} $  and $k_{p0}$ ($\omega_{p1} $ and $k_{p1}$) as the wave frequency  and vector of the laser in the laboratory frame (co-moving frame). 
It is usually the case that $k_{p1} > k_{pp1}$ and 
define $ S = k_{p1} / k_{pp1} > 1$.  Then,   
using Eq.~(\ref{eq:lorentz1}), we obtain the relationship of the laser frequencies  (vectors)
between the laboratory frame and co-moving frame; 

\begin{eqnarray} 
\frac{\omega_{p0}}{\sqrt{\gamma_0} \omega_{pe} }
 &=& \left( \sqrt{1 + 3 S^2 } - \sqrt{3} \beta S  \right)  = \Delta_s \mathrm{,} \nonumber \\ \nonumber \\ 
 \omega_{p1} &=& \sqrt{1+ 3S^2} \left(\frac{\omega_{pe} }{\sqrt{\gamma_0}}\right) \mathrm{.}
\label{eq:pump}  
\end{eqnarray} 
From Eq.~(\ref{eq:pump}),
Eq.~(\ref{eq:S}) can be derived  assuming $\beta \cong 1$.
The BRS scattering of the laser and the plasmons excited by the first laser is given in the co-moving frame  by~\cite{McKinstrie}:
\begin{equation}
\left( \frac{\partial }{\partial t} + v_s \frac{\partial}{\partial x} + \nu_2\right)A_s  = -ic_s A_p A^*_3   \label{eq:2} \mathrm{,}
\end{equation}
where $A_i= eE_{i1}/m_e\omega_{i1}c$  is 
the ratio of  the electron quiver velocity of the pump pulse ($i=p$)
and seed pulse ($i=s$) relative to the velocity of the light $c$, 
 $A_3 = \tilde{n}/n_1$ is the the Langmuir wave amplitude,
$\nu_2$ is the rate of the inverse bremsstrahlung  
of the seed, 
$ c_2 = \omega_3^2/ 2 \omega_{p1}$, and $\omega_3 \cong \omega_{pe} / \sqrt{\gamma_0} $ is 
the plasmon frequency in the co-moving frame. 
The energy and momentum conservation of the BRS is given as 
\begin{eqnarray} 
 \omega_{p1} &=& \omega_{s1} + \omega_{3} 
\mathrm{,} \nonumber \\  
  k_{p1}&=& k_{s1}+k_3 \mathrm{,} \label{eq:cons} 
\end{eqnarray}
where $k_3$ is the wave vector of the plasmon.
Define $P = k_{s1} / k_{pp1} $  and  we obtain 
   $P \cong S - 1/\sqrt{3} $
 from the energy conservation of Eq.~(\ref{eq:cons}), 
$ \sqrt{1 + 3 S^2 }= \sqrt{1+3P^2}+1$. 
The frequency of the seed pulse is given from the Lorentz transform as 
 \begin{equation}
\frac{\omega_{s0}}{\sqrt{\gamma_0} \omega_{pe} } 
 = \left( \sqrt{1 +P^2 } + \sqrt{3} \beta P \right)  = \Delta_p \label{eq:seed} \mathrm{.}
\end{equation}
Using Eqs.~(\ref{eq:pump}) and (\ref{eq:seed}),
the ratio between the plasmon pump frequency and the seed pulse (soft x-ray) in Eq.~(\ref{eq:down2}) can be derived.

\section{mean-free path of the BRS and conversion efficiency} 

From Eq.~(\ref{eq:2}),
the considerable part of the pump energy will be transferred to the 
seed pulse when $c_s A_3 \delta t_b \cong 1$, where $t_b$ is the BRS interaction time in the co-moving frame and we obtain the mean-free path 
\begin{equation} 
l_b = \delta t_b c  \cong c (2 \omega_{s1}  /\omega_3^2 ) (1/A_3) 
\mathrm{.} \label{eq:mean}
\end{equation}
 On the other hand, 
the Thomson scattering suggests that  $ l_t \cong 1/n\sigma_t $ with $\sigma_t = (mc^2 / e^2)^2 $.  For an example, when $n_1 \cong 10^{20} / \mathrm{cc} $,   
we estimate $l_t \cong 10^3 \ \mathrm{cm} $ and $l_b \cong (10^{-4} / A_3) S  \ \mathrm{cm}$.  Even for $A_3 \cong 0.001 $, the soft x-ray  radiation by the BRS is considerably stronger than the Thomson scattering or $l_t \gg l_b $

The maximum conversion efficiency from the pump energy to the seed energy can be estimated as follows. 
Denote the total energy of the pump laser (the seed laser) in the laboratory frame
 as $\mathrm{E}_{p0} $ ($\mathrm{E}_{s0}$).  
In the co-moving frame, 
the pump energy is seen to be as 
$\mathrm{E}_{p1}  \cong (\sqrt{3} S\mathrm{E}_{p0}/ 
\Delta_s \gamma_0 )$ 
from Eq.~(\ref{eq:pump}). 
Considering the conversion efficiency in this co-moving as $\epsilon_1 $, 
the energy of the seed pulse is given as 
  $\mathrm{E}_{s1} =  \epsilon_1 (\sqrt{3} S\mathrm{E}_{p0}/ 
\Delta_s \gamma_0 ) $. 
This energy of the seed pulse is seen in the laboratory to be 
   $\mathrm{E}_{s0} =  
\Delta_p \gamma_0 \mathrm{E}_{s1}  / \sqrt{3}S =  \epsilon_1 (\Delta_p / \Delta_s) \mathrm{E}_{p0}$. 
Then, the conversion efficiency in the laboratory frame is given as 

\begin{equation} 
\epsilon_0 =    \left(\frac{\Delta_p }{\Delta_s} \right) \epsilon_1  
\cong  \left( \frac{ 
\sqrt{(1 + 3 P^2)} + \sqrt{3} \beta P }{ 
{\sqrt{(1 - 3 S^{2})} -\sqrt{3} \beta S }}\right) \epsilon_1  \mathrm{.}
\label{eq:conv} 
\end{equation}
The estimation of $\epsilon_1$ in the co-moving frame follows.
We consider two cases;  
 when the Langmuir waves are rather excited isotropically  and 
 when they are excited preferentially in the beam direction. 
If the plasmons are excited isotropically, 
the BRS would be radiated isotropically.
However, only the radiations in the direction of the beam  are
 relevant for the soft-x ray since 
only those radiations would be up-shifted 
by the Doppler effect in the laboratory frame; 
The angular width relevant for the soft x-ray would be given as $d\theta  \cong S/ \gamma_0 \Delta_p $  and 
the relevant portion of the radiation is then $(d \theta)^2$ or $\epsilon_1 \cong (S / \gamma_0 \Delta_p )^2$. 
In the case when the plasmon distribution is sharply peaked at $\theta \cong 0$, 
the most of the radiation would be in the direction of the beam and $\epsilon_1 \cong 1$. 
From the above consideration, we could estimate the optimal conversion efficiency as 

\begin{equation}
 \left(\frac{S^2}{\Delta^2_p \gamma_0^2 }\right)\left(\frac{\Delta_p}{ \Delta_s }\right) <  \epsilon_0  <  \frac{\Delta_p}{ \Delta_s } \mathrm{.}  \label{eq:eps}
\end{equation}
Eq.~(\ref{eq:eps}) is the maximum possible efficiency since we assume that most of the BRS pump energy will be radiated via the BRS. 
Given the fact that $\epsilon_1$ could  be a few tens of percents in an optimistically envisioned scenario, it is possible that $\epsilon_0> 1 $ 
or the conversion efficiency could be larger than 100 \%. 
This apparent contradiction can be understood by estimating 
the energy and momentum conservation of the BRS in the laboratory frame. 
In the co-moving frame, 
the BRS pump laser and plasmons will be scattered into a seed photon. 
In this process, 
the laser photon and plasmon gives the energy and momentum 
to the seed photon; the plasmon (the BRS pump laser) gives  the momentum $-k_{s1} - k_{p1} $ ($k_{pl} $)  
and the seed photon acquires $-k_{s1} $ momentum. 
In the laboratory frame, 
the momentum (energy) of the plasmon is much larger than the BRS pump photon due to the Doppler effect and 
the seed pulse obtains  most of its energy not 
from the pump pulse 
but from the plasmon (by the ratio of $1:\Delta_P/\Delta_s$). 
It is then not contradictory that $\epsilon_0 $ can be larger than the unit. 
The major energy source of the seed is not the BRS pump laser 
but the plasmons excited by the plasmon pump laser, 
which is contrary to the conventional BRS 
compression~\cite{Fisch, malkin1, sonbackward}. 
The energy of the plasmon comes from the electron beam kinetic energy.

As an example, we provide the estimations for  a few physical  parameters 
for the electron beam and the plasmon and BRS pump laser in Table~(\ref{tb}). 
In the table, 
we consider the plasmon pump intensity of $I= 10^{11}  \ \mathrm{W} / \sec$ and the electron beam of the 5 keV electron temperature  in the co-moving frame. 
From the  examples,  
we conclude that, for the most effective soft-xray generations, 
the $S$ ($\gamma_0$) should be in the range $1<S<5 $ ($2<\gamma_0<10$). 
The most effective way for the two-plasmon decay seems to be case when 
the wave vector of the plasmon pump laser is in the same direction with the electron beam. The wave length of the soft x-ray generated ranges from $ 5 \ \mathrm{nm} $ to  $300  \ \mathrm{nm} $. However, 
the most practical radiation from this scheme seems to be 
the soft x-ray from $10 \ \mathrm{nm} $ to  $50  \ \mathrm{nm} $. 
The conversion efficiency could be as high as a few thousand percents and the BRS wave length is always shorter than the Thomson scattering in those examples.


\begin{table}[t]
\centering
\begin{tabular}{|c||cccccc||}
	\hline
   &  1    &   2  & 3 & 4  & 5 & 6  \\
	\hline \hline 
  $n_{16}$  &  20.6 &  3.78 &    0.28 &        0.57 &    10.0&   19.2  \\  
 $\gamma_0$  & 3  &  3&    3 &       10 & 15 &   7 \\
 $S$         &4&   10&    3&       3&    15&    5  \\ 
$\lambda_{pp0}$ &  1.15 &  2.7&      9.8 &          5.1 &     1.0& 1.0  \\
$\lambda_{p0} $  & 1.0 &   1.0 &     10.0&      10.0&    3.6 &    2.0 \\ 
$\mathbf{\lambda_{s0}}$  & \textbf{ 31.4} &  \textbf{29.4} &    \textbf{353.7} &     \textbf{133.5} &   \textbf{5.2}&      \textbf{16.6} \\
$A_3$  &  0.022    & 0.128 &    0.14 &      0.07 &      0.2&     0.076 \\
$l_t$  & 436 &  2380 &    31399 &     52426 &      4483 &       1089 \\
$l_b$ & 0.005 &     0.005 &       0.005 &     0.013 &       0.006 &      0.003 \\ 
$\Delta_p / \Delta_s$ & 32 &    34.9 &      28.5 &    75.5 &       687 &   120 \\ 
$d\theta$  & 0.098 &   0.09 &       0.09 &    0.029 &       0.02 &    0.04  \\
\hline
\end{tabular}
\caption{Table 1: the laser and electron beam parameter and the 
characteristic of the THz radiation. \label{tb}
$n_{20} $ is the electron density $n_0$ normalized by $10^{20} / 
\mathrm{cc} $,  $S $ is defined in Eq.~(\ref{eq:S}), $\lambda_{pp1} $ and $\lambda_{p0} $ is the plasmon and BRS pump laser normalized by $ 1  \ \mu \mathrm{m}$, 
 $\lambda_{s0} = 2 \pi c/  \omega_{s0} $ is the wave length of the seed pulse in the unit of nm from Eq.~(\ref{eq:down2}),  $A_3=\delta n / n_1$ is the plasmon intensity given in Eq.~(\ref{eq:lan}), $l_t$ ($l_b$) is the mean-free path of the Thomson scattering  (BRS) in the unit of cm from Eq.~(\ref{eq:mean}), 
$\Delta_p / \Delta_s $ is the ratio of the seed pulse frequency to the BRS pump pulse frequency as given in Eq.~(\ref{eq:seed}), and 
$d \theta = S/\Delta_P \gamma_0 $ is the anglur width in the conversion efficiency estimation.  
}
\end{table}

\section{summary}

In summary, we propose a scheme for soft x-ray radiations. The scheme 
is based on the laser-plasm interaction 
in a moderately relativistic electron beam; namely the two-plasmon decay 
and the backward Raman scattering. 
In the scheme, the first laser (the plasmon pump) 
excites the plasmons inside the electron beam via the two-plasmon decay~\cite{liu, langdon}. 
There are three possibilities for the two-plamson decay and those three cases are compared.
The most prominent one is to excite the plasmons in the same direction to the beam. 
The second laser (the BRS pump) emits the xoft x-ray via the BRS with the excited plasmon. 
The ratio of the seed pulse frequency  to the BRS pump frequency 
 is given in Eq.~(\ref{eq:seed}).
We  estimate the conversion efficiency from the BRS pump to the soft x-ray and conclude that it is  sometimes larger than the unit.  This is not contradictory since 
 the energy for the soft x-ray is mostly not from the BRS pump laser but from the plasmon energy. 
Using the plasmon pump and the BRS pump with the wave length of 1 $\mu$m to 10 $\mu$m and intensity of $I\cong 10^{12}\  \mathrm{W} / \mathrm{cm}^2$ and 
the electron beam with  density of $10^{18} / \mathrm{cc} $ to $10^{22} 
/ \mathrm{cc} $  and the relativistic factor of $ 1 < \gamma_0 < 100 $, 
 the scheme can generate the soft x-ray from  10 nm to 300 nm with the duration of  10 femto seconds to 1 pico-second. 


This scheme has a few drawbacks. First, it needs dense and uniform electron beam
 for efficient two-plasmon decay. 
Second, once the electron beam density and the relativistic factor are fixed, 
there is no freedom in the frequency for the plasmon pump laser as given in
Eq.~(\ref{eq:ppump});
 The auto tuning is harder than other schemes. 
Third, two lasers instead of one are needed. 
Fourth, because the electron beam density is high,  it needs to propagate inside the plasma of comparable density in order to avoid the space charge expansion. 
However, with all these drawbacks, 
given the low intensity threshold of the plasmon pump laser and the possible high conversion efficiency, 
this scheme is very attractive as an alternative soft x-ray source. 
In the two-plasmon decay, the saturation mechanism is the ion-dynamics~\cite{langdon}. 
But in the current scheme, the ion might have relativistic velocity in the co-moving frame and its effective mas could be higher. Then, the non-linear saturation mechanism in the conventional two-plasmon decay~\cite{langdon} might be further mitigated leading to the stronger BRS.
\bibliography{tera2}

\end{document}